\documentclass[twocolumn,prl,superscriptaddress,showpacs,byrevtex]{revtex4}
\usepackage{epsf,graphicx,amssymb}

\begin{document}
\title{Observation of gravity-capillary wave turbulence}
\author{\'Eric Falcon}
\altaffiliation[Permanent address: ]{Mati\`ere et Syst\`emes Complexes, Universit\'e Paris 7, CNRS UMR 7057, France}
\author{Claude Laroche}
\affiliation{Laboratoire de Physique, \'Ecole Normale Sup\'erieure de Lyon, UMR 5672, 46, all\'ee d'Italie, 69 007 Lyon, France}
\author{St\'ephan Fauve}
\affiliation{Laboratoire de Physique Statistique, \'Ecole Normale Sup\'erieure, UMR 8550, 24, rue Lhomond, 75 005 Paris, France}

\date{\today}

\begin{abstract}  
We report the observation of the cross-over between gravity and capillary wave turbulence on the surface of mercury. The probability density functions of the turbulent wave height are found to be asymmetric and thus non Gaussian. The surface wave height displays power-law spectra in both regimes. In the capillary region, the exponent is in fair agreement with weak turbulence theory. In the gravity region, it depends on the forcing parameters. This can be related to the finite size of the container. In addition, the scaling of those spectra with the mean energy flux is found in disagreement with weak turbulence theory for both regimes.

\end{abstract}

\maketitle

Wave turbulence, also known as weak turbulence, is observed in various situations: internal waves in the ocean \cite{Lvov04}, surface waves on a stormy sea \cite{Toba73}, Alfv\'en waves in astrophysical plasmas \cite{Sagdeev79}, Langmuir waves \cite{Huba80} and ion waves \cite{Mizuno83} in plasmas, spin waves in solids. It has been also emphasized that wave turbulence should play an important role in nonlinear optics \cite{Kuznetsov91}. However,  wave turbulence experiments are scarce.
Most of them concern capillary or gravity waves. For short wavelengths,  capillary wave turbulence has been observed by optical techniques \cite{Wright96,Holt96,Lommer02,Brazhnikov02}. It has been reported that the height of the surface displays a power-law frequency spectrum $f^{-17/6}$ in agreement with  weak turbulence (WT) theory \cite{Zakharov67Cap} and simulations \cite{Pushkarev96}. For longer wavelengths,  gravity wave turbulence has been mainly observed {\it in situ} (i.e. on the sea surface or in very large tanks) with wind-generated waves leading to power-law spectra $f^{-4}$ \cite{Toba73} in agreement with isotropic WT theory \cite{Zakharov67Grav} and simulations \cite{Onorato02}. However, when the turbulence is not forced by wind or by an isotropic forcing, mechanisms of energy cascade in the inertial regime change, as well as the scaling law of the spectrum \cite{Onorato02,Kitaigorodskii83,Kuznetsov04}, 
and are still a matter of debate. 

Besides scalings with respect to frequency or wave number, Kolmogorov-type spectra also depend on the mean energy flux $\epsilon$ cascading from injection to dissipation. This dependence is related to the nature of nonlinear wave interactions which are different in capillary (3-wave interactions) versus gravity (4-wave interactions) regimes \cite{Zakharov67Cap,Zakharov67Grav}. To our knowledge, the mean energy flux has never been measured in wave turbulence and no experiment has been performed to study how spectra scale with $\epsilon$. Matching the gravity and capillary spectra and WT theory breakdown are other open questions \cite{Newell92,Newell03}. 
We report in this letter how power-law spectra in the gravity and capillary ranges depend on the forcing parameters of surface waves. We measure the mean energy flux $\epsilon$ and show that, although the scaling of the spectra with respect to frequency looks in agreement with WT theory in some limits, their scaling on $\epsilon$ differ from theoretical predictions. 

\begin{figure}[h]
\centerline{
\epsfysize=65mm
\epsffile{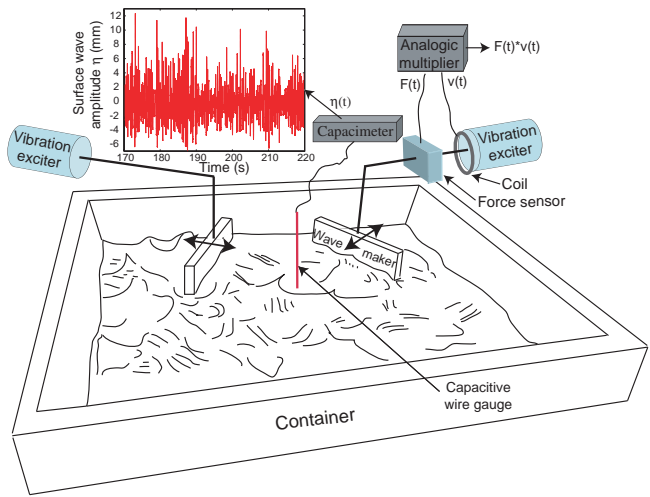} 
}
\caption{(color online). Schematic view of the experimental setup showing a typical time recording of the surface wave height, $\eta(t)$, at a given location during 50 s. $\langle \eta \rangle \simeq 0$.} 
\label{fig01}
\end{figure}

The experimental setup consists of a square plastic vessel, 20 cm side, filled with mercury up to a height, $h$ ($h=18 $ mm in most experiments) (see Fig.\ \ref{fig01}).  The properties of the fluid are, density, $\rho = 13.5$ 10$^{3}$ kg/m$^3$, kinematic viscosity, $\nu=1.15$ 10$^{-7}$ m$^2$/s and surface tension $\gamma=0.4$ N/m. Contrary to the usual bulk excitation of waves by Faraday vibrations \cite{Wright96,Lommer02}, surface waves are generated by the horizontal motion of two rectangular ($10\times 3.5$ cm$^2$) plunging PMMA wave makers driven by two electromagnetic vibration exciters (BK 4809) via a power supplied (Kepco Bop50-4A).  The wave makers are driven with random noise excitation, supplied by a function generator (SR-DS345), and selected in a frequency range 0 - $f_{driv}$ with  $f_{driv}=4$ to 6 Hz by a low-pass filter (SR 640). This corresponds to wavelengths of surface waves larger than $4$ cm. This is in contrast with most previous experiments on capillary wave turbulence driven by one excitation frequency \cite{Brazhnikov02,Wright96,Lommer02}. Surface waves are generated $2.2$ cm inward from two adjacent vessel walls and the local displacement of the fluid in response to these excitations is measured $7$ cm away from the wave makers. 
A capacitive wire gauge, perpendicular to the fluid surface at rest, is made of an insulated copper wire, 0.1 mm in diameter. The insulation (a varnish) is then the dielectric of an annular capacitor with the wire as the inner conductor and mercury as the outer one. The capacitance is thus proportional to the fluid level. A low-cost home-made analogic multivibrator with a response time 0.1 ms is used as a capacitance meter in the range 0 -- 200 pF. The linear sensing range of the sensor allows wave height measurements from 10 $\mu$m up to 2 cm with a 20 mm/V sensitivity.
Although resistance or capacitance wire probes are widely used to get precise measurements of the level of quasi-static liquids, their dynamical response in the case of a rapidly varying wavy surface is not well known due to possible meniscus effects \cite{Lange82}. Thus, we have first checked our results with measurements performed with eddy current displacement transducers or with an optical determination of the local slope of the surface \cite{Falcon}. 

The mean energy flux injected by the wave makers and dissipated by viscosity is determined as follows. The velocity $V(t)$ of the wave maker is measured using a coil placed on the top of the vibration exciter (see Fig.\ \ref{fig01}). The {\it e.m.f.} generated by the moving permanent magnet of the vibration exciter is proportional to the  excitation velocity. For a given excitation band width, the $rms$ value, $\sigma_V$, of the velocity fluctuations of the wave maker is proportional to the driving voltage $U_{rms}$ applied to the vibration exciter. The force $F(t)$ applied by the vibration exciter to the wave maker is measured by a piezoresistive force transducer (FGP $10$ daN). The power injected into the fluid by the wave maker is $I(t) = -F_R (t) V(t)$ where $F_R (t)$ is the force applied by the fluid on the wave maker. It generally differs from $F(t) V(t)$ which is measured here because of the piston inertia. However, their time averages are equal, thus $\langle I \rangle = \langle F(t) V(t) \rangle$. \\

A typical recording of the surface wave amplitude at a given location is displayed in the inset of Fig.\ \ref{fig01} as a function of time. The wave amplitude is very erratic with a large distribution of amplitudes. The largest values of the amplitude are of the order of the fluid depth, whereas the mean value of the amplitude is close to zero.

\begin{figure}[h]
\centerline{
\epsfysize=65mm
\epsffile{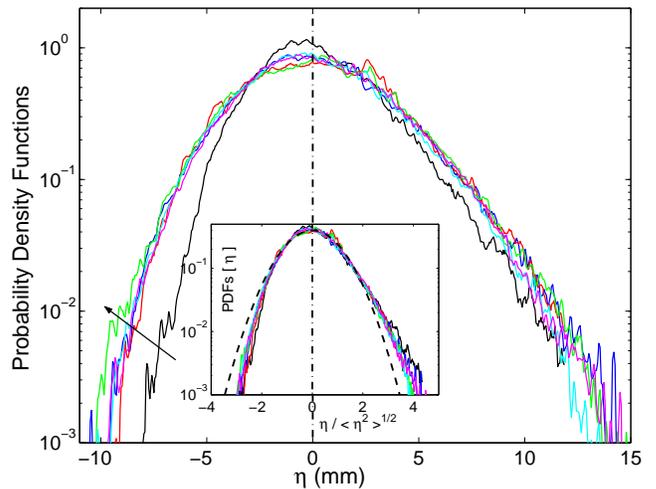} 
}
\caption{(color online). Probability density functions of the wave-height, $\eta$, for the maximum excitation amplitude ($U_{rms}= 0.9$ V) and for 6 different values of the fluid depth, from $h= 18$, $35$, $55$, $80$, $110$ to $140$ mm (see the arrow). The frequency band is $0 \leq f \leq 6$ Hz. Inset: Same PDFs displayed using the reduced variable $\eta / \sqrt{\langle \eta^2 \rangle}$. Gaussian fit with zero mean and unit standard deviation ($--$).  }
\label{fig02}
\end{figure}

The probability density function (PDF) of the surface wave height, $\eta$, is found to be Gaussian at low forcing amplitude (not shown here), whereas it becomes asymmetric at high enough forcing (see Fig.\ \ref{fig02}). The positive rare events such as high crest waves are more probable than deep trough waves \cite{ruban06}. This can also be directly observed on the temporal signal $\eta(t)$ shown in the inset of Fig.\ \ref{fig01}.  A similar asymetrical distribution is observed when using water instead of mercury, although the meniscus has an opposite concavity. As shown in Fig.\ \ref{fig02}, the asymmetry is enlarged when the largest trough to crest amplitudes become comparable to the height of the layer. However, it persists in the limit of deep water waves.
Note that the mean value $\langle \eta \rangle$ remains close to zero and the PDFs  of the reduced variable $\eta / \langle \eta^2 \rangle^{1/2}$ roughly collapse (see inset of Fig.\ \ref{fig02}).

\begin{figure}[h]
\centerline{
\epsfysize=65mm
\epsffile{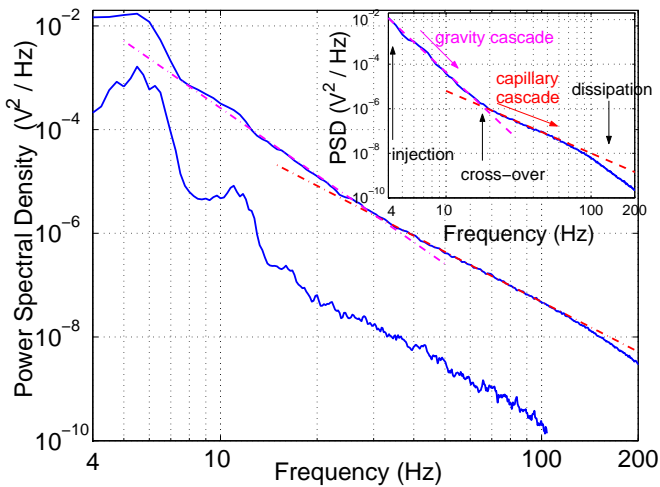} 
}
\caption{(color online). Power spectra of the surface wave height for two different driving voltages $U_{rms}=0.2$ and $0.9$ V (from bottom to top). The frequency band is $0 \leq f \leq 6$ Hz.  Dashed lines have slopes -4.3 and -3.2.  Inset: The frequency band is $0 \leq f \leq 4$ Hz, and $U_{rms} = 0.9$ V. Dashed lines had slopes of -6.1 and -2.8.}
\label{fig03}
\end{figure}

The power spectrum of the surface wave amplitude is recorded from 4 Hz up to 200 Hz and averaged during 2000 s. For small forcing, peaks related to the forcing and its harmonic are visible in the low frequency part of the spectrum in Fig.\ \ref{fig03}.  At higher forcing, those peaks are smeared out and a power-law can be fitted. At higher frequencies, the slope of the spectrum changes, and a cross-over is observed near 30 Hz between two regimes. This corresponds to the transition from gravity to capillary wave turbulence. At still higher frequencies (greater than 150 Hz), viscous dissipation dominates and ends the  energy cascade. For a narrower frequency band of excitation (0 - 4 Hz), similar spectra are found but with a broader power-law in the gravity range (see inset of Fig.\ \ref{fig03}). When the two wave makers are driven with two noises with different band widths, {\it e.g.} 0 - 4 Hz and 0 - 6 Hz, the harmonic peak is no longer present, and gravity spectra display a power-law even at low driving amplitude. 
 
\begin{figure}[h]
\centerline{
\epsfysize=65mm
\epsffile{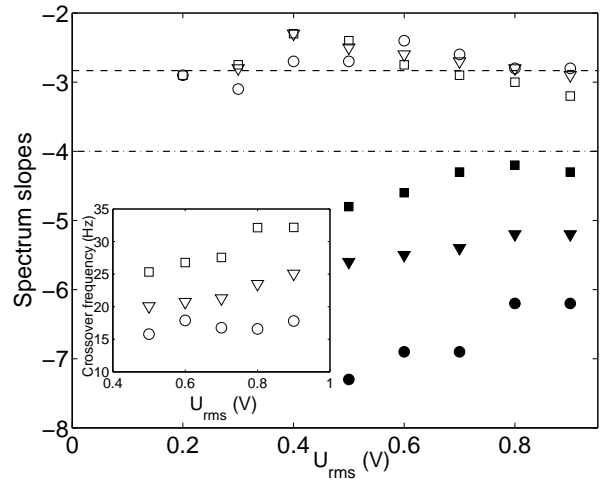} 
}
\caption{Slopes of surface-height spectra for gravity (full symbols) and capillary (open symbols) waves for different forcing band widths and intensities: ($\circ$) 0 to 4 Hz, ($\bigtriangledown$) 0 to 5 Hz and ($\Box$) 0 to 6 Hz. Power-law exponents of gravity wave spectrum ($-\cdot$) and capillary waves spectrum ($--$) as predicted by WT theory
(Eq. \ref{WKT}). Inset: Cross-over frequency between gravity and capillary regimes as a function of the forcing intensity and band width.}
\label{fig04}
\end{figure}

For linear waves, the cross-over between gravity and capillary regimes corresponds to  a wave number $k$ of the order of the inverse of the capillary length $l_c \equiv \sqrt{\gamma/\left(\rho g\right)}$, {\it i.e.} to a critical frequency, $f_c = \sqrt{g/2 l_c}/{\pi}$, where $g$ is the acceleration of gravity. For mercury, $l_c=1.74$ mm and $f_c\simeq17$ Hz corresponding to a wavelength of the order of 1 cm. The insets of Fig.\ \ref{fig03} and Fig.\ \ref{fig04} show a correct agreement in the case of a narrow driving frequency band. We also observe that the cross-over frequency increases with the driving amplitude and with the width of the driving frequency band (see the inset of Fig.\ \ref{fig04}). 
This can be due to the fact that the above estimate of $f_c$ is only valid for linear waves. 
The capillary length cannot be significantly changed using other interfaces between simple liquids and air. It is at an intermediate scale between the size of the experiment and the dissipative length. In this laboratory-scale experiment, this limits both the gravity and capillary regimes to less than a decade in frequency. With laboratory-scale experiments, we can study full range gravity waves with a liquid-vapor interface close to its critical point and full range capillary waves in a micro-gravity environment.

Surface wave turbulence is usually described as a continuum of interacting waves governed by kinetic-like equations in case of small nonlinearity and weak wave interactions.  WT theory predicts that  the surface height spectrum $S_{\eta}(f)$, i.e. the Fourier transform of the autocorrelation function of $\eta (t)$,  is scale invariant with a power-law frequency dependence. Such a Kolmogorov-like spectrum writes
\begin{equation}
\begin{array}{ll}
S_{\eta}(f) \propto \epsilon^{\frac{1}{2}} \left(\frac{\gamma}{\rho}\right)^{\frac{1}{6}} f^{-\frac{17}{6}} {\rm \ for\  capillary\ waves\ [11]} ,\\  
S_{\eta}(f) \propto  \epsilon^{\frac{1}{3}} g f^{-4} {\rm \ for\ gravity\ waves\ [13]}, 
\end{array}
\label{WKT}
\end{equation}
where $ \epsilon$ is the energy flux per unit surface and density [$S_{\eta}(f)$ has dimension $L^2T$ and $\epsilon$ has dimension $(L/T)^3$].
In both regimes, these frequency power-law exponents are compared in Fig.\ \ref{fig04} with the slopes of surface height spectra measured for different forcing intensities and band widths. The experimental values of the scaling exponent of capillary spectra are close to the expected $f^{-2.8}$ scaling as already shown with one driving frequency \cite{Wright96,Lommer02,Brazhnikov02} or with noise \cite{Brazhnikov02}. Figure\ \ref{fig04} shows that this exponent does not depend on the amplitude and the frequency band of the forcing, within our experimental precision. For the gravity spectrum, no power-law is observed at small forcing since turbulence is not strong enough to hide the first harmonic of the forcing (see Fig.\ \ref{fig03}). At high enough forcing, the scaling exponent of gravity spectra is found to increase with the intensity and the frequency band (see Fig.\ \ref{fig04}). For gravity waves, the predicted $f^{-4}$ scaling of Eq.\ (\ref{WKT}) is only observed for the largest forcing intensities and band width (see  Fig.\ \ref{fig04}). The dependence of the slope of the gravity waves spectrum on the forcing characteristics can be ascribed to finite size effects \cite{finitesize}. Similar results in the gravity range have been recently found in a much larger tank with sinusoidal forcing \cite{denissenko06}.\\

\begin{figure}[h]
\centerline{
\epsfysize=65mm
\epsffile{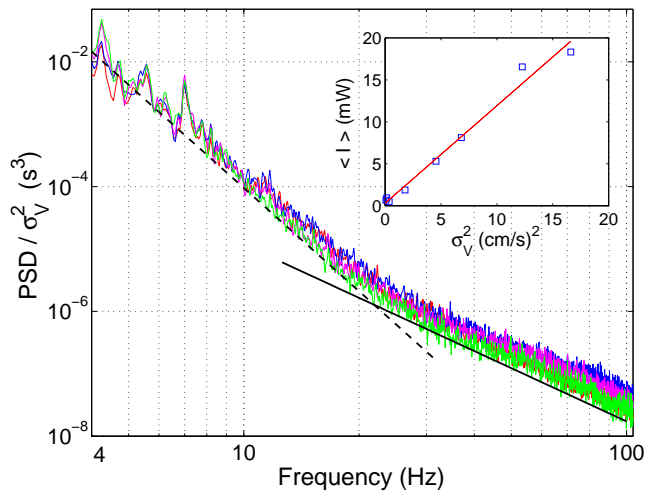} 
}
\caption{(color online). Spectra of the surface wave amplitude divided by the variance $\sigma_V^2$ of the velocity of the wave maker for different forcing amplitudes, $\sigma_V = 2.1, 2.6, 3.5$ and $4.1$ cm/s. The frequency band is $0 \leq f \leq 4$ Hz. The dashed line has slope $-5.5$ whereas the full line has slope $-17/6$. The mean injected power is displayed as a function of  $\sigma_V^2$ in the inset. The best fit gives a slope $11.5$ W/(m/s$^2$).
}
\label{fig05}
\end{figure}

We finally consider how those spectra scale with the mean energy flux $\epsilon \equiv \langle I \rangle / (\rho S_P)$ where $\langle I \rangle$ is the mean power injected by the wave maker and $S_P$ is the area of the wave maker. With given $\sigma_V$, we have first check that $\langle I \rangle$ in proportional to $S_P$ and decreases by a factor $13$ when mercury is replaced by water. Our measurements also show that $\langle I \rangle \propto \sigma_V^2$ with a proportionality coefficient of order $10$ W/(m/s)$^2$ (see the inset of Fig.\ \ref{fig05}). We thus have $\epsilon \propto c \sigma_V^2$ where $c$ has the dimension of a velocity. 
If we assume that $\epsilon$ should involve only large scale quantities, it cannot depend on surface tension or viscosity. Then $c$ is a characteristic gravity wave speed at large wave length. 
The dependence of $\epsilon$ on $c$ can be ascribed to finite size effects. The inverse travel time of a wave within the tank or the frequency difference between the discrete modes of the tank both scale with $c$. Discreteness also explains why a spectrum with enlarged peaks is observed in Fig.\ \ref{fig04} at low forcing in the gravity range. However, the large enough values of $\epsilon$ required to observe power laws, are more than one order of magnitude smaller than the critical flux  $(\gamma g / \rho)^{3/4} \approx 2200$ (cm/s)$^3$ corresponding to the breakdown of weak turbulence \cite{Newell03}.

The best choice in order to collapse our experimental spectra on a single curve for different values of $\sigma_V$ is displayed in Fig.\ \ref{fig05} where the power spectral density divided by  $\sigma_V^2$ is plotted versus $f$. Surprisingly, spectra are collapsed on both the gravity and capillary ranges by this single scaling. Their dependence on the mean energy flux $\epsilon$ thus corresponds neither to prediction of WT theory for the capillary regime ($\epsilon^{1/2}$) nor to the one related to the gravity regime ($\epsilon^{1/3}$) but is linear with $\epsilon$. 
This discrepancy can result from several reasons.
First, the size of the container is too small to reach a forcing-independent gravity regime. Second, capillary and gravity regimes probably interact such that it may be wrong to consider them independently as in Eq. (\ref{WKT}). Third, we observed that the energy flux strongly fluctuates and takes both positive and negative instantaneous values much larger than its mean. The possible effect of these fluctuations on wave turbulence deserves further studies.

\begin{acknowledgments}
We thank F. Palierne for his help and B. Castaing, L. Biven, S. Nazarenko and A. Newell  for fruitful discussions. This work has been supported by the French Ministry of Research (ACI 2001) and by the CNES.\\
\end{acknowledgments}


\end{document}